\documentclass[aps,12pt,showpacs]{revtex4}

\usepackage{graphicx}
\usepackage{dcolumn}
\usepackage{amssymb,amsmath}
\usepackage{natbib}

\def\be{\begin{equation}}
\def\ee{\end{equation}}
\def\bea{\begin{eqnarray}}
\def\eea{\end{eqnarray}}

\def\begineq{\begin{equation}}
\def\endeq{\end{equation}}

\begin{document}

\title{Logarithmic temperature profiles in the ultimate regime of thermal convection}
\author{Siegfried Grossmann$^{1}$  and Detlef Lohse$^{2}$}
\affiliation{$^1$ Fachbereich Physik, Renthof 6, Philipps-Universitaet Marburg, D-35032 Marburg, Germany\\
$^{2}$ Department of Physics, Impact Institute,  and J.\ M.\ Burgers Centre for
Fluid Dynamics,\\ University of Twente, 7500 AE Enschede, The Netherlands}

\date{\today}

\begin{abstract}
We report on the theory of logarithmic temperature profiles in very strongly developed thermal convection in the geometry of a Rayleigh-B\'enard cell with aspect ratio $\Gamma = 1$ and discuss the degree of agreement with the recently measured profiles in the ultimate state of very large Rayleigh number flow. The parameters of the log-profile are calculated and compared with the measure ones. Their physical interpretation as well as their dependence on the radial position are discussed.       
\end{abstract}

\maketitle

\section{Introduction}

The ultimate state of thermal convection in a Rayleigh-B\'enard cell is generally believed to be fully turbulent, both in the bulk and in the boundary layers and 
with respect to both the velocity as well as to the temperature fields. As shown in ref.\ \cite{gro11}, this implies very characteristic scaling behavior of the heat transport flux, $Nu \propto Ra^{0.38}$ as well as of the strength of the large scale flow, $Re \propto Ra^{0.50}$. Here the Rayleigh number $Ra$, the Nusselt number $Nu$, and the Reynolds number $Re$ are defined as usual, see e.g.\ \cite{ahl09,loh10}. In a series of recent experiments, cf.\ \cite{fun09, ahl09b, ahl10, he12a, ahl12a}, these scaling laws of $Nu$ and $Re$ have been measured and reported. 

In the previous work \cite{gro11} we have calculated the {\it global} scaling exponents in the ultimate state.
 This ultimate state theory, though derived by employing the characteristic profiles of fully developed turbulent flows, concentrated on the implications for and the interpretation of the various global scaling exponents. We did not explicitly describe and report the {\it local} profiles themselves.  
Meanwhile the local thermal profiles in the very large $Ra$-regime have been measured, cf.\ \cite{ahl12}. We thus present here the corresponding 
local profiles, in the spirit of and in extension of our earlier theory from ref.\ 
 \cite{gro11}. 

\section{Thermal log-profiles of flow along plates}

We start from the equations of motion for the thermal field $T(\vec{x},t)$. The involved velocity field is understood to solve the corresponding Navier-Stokes equation for an incompressible flow, $\nabla \cdot \vec{u} = 0$, i.e., we assume that the temperature behaves like a passive scalar.
 The molecular properties kinematic viscosity $\nu$ and thermal diffusivity $\kappa$ are taken as temperature independent fluid parameters. It is 
\be
\partial_t T = - \vec{u} \cdot \vec{\nabla} ~T + \kappa \Delta T,
\label{T-equation}
\ee   
\noindent together with the respective boundary conditions (b.c.) $T(z=0) = 0$ and $T(z=L) = - \Delta$ and no heat flux at the sidewalls. In order to understand its turbulent solution we model the RB flow in a rather simplified way, which in the first steps allows to make some use of textbook wisdom, cf. \cite{ll87} \S 54, or \cite{pop00} \S 7.1.4. Consider a flow along an infinite plate in $x$-direction and ask for the profile in $z$-direction perpendicular to the plate. First decompose the fields into their long time means $\vec{U}, \Theta$ and their fluctuating parts $\vec{u}', \theta'$. The time averages depend on $z$ only and the velocity has an $x$-component only, choosen to be in positive $x$-direction. Per definition we choose $\Theta(z=0) = 0$. If there is heat flow upwards, $J = \langle u_3\theta\rangle_{A,t} - \kappa \partial_3 \langle \theta \rangle_{A,t}$ positive, $\Theta$ decreases with increasing distance $z$ from the plate and vice versa. This describes the !
 situation at the bottom plate of an RB-cell. - At the top plate the temperature flux $J$ is still positive, directed towards the plate, but $\Theta$ increases with growing distance from the top plate or decreasing $z$, starting from $z=L$ downwards. 
\be
0 = -\vec{U}\cdot \vec{\nabla} ~\Theta - \overline{\vec{u'} \cdot \vec{\nabla} \theta'}^t + \kappa \Delta ~ \Theta.
\label{timeaveragedTequation}
\ee     
The first term vanishes, since only the $x$-derivative can contribute but $\Theta$ depends on z only. The last term for the same reason reads $\kappa \partial_z^2 \Theta(z)$. The second term has to be modeled; we use the mixing length type Ansatz for an eddy thermal diffusivity. First, $\overline{u_i' \theta'}^t$ as a long time mean depends on $z$ only, is approximated by employing the temperature gradient hypothesis, 
\be
\overline{u_i'\theta'}^t \hat{=} - \kappa_{turb}(z) ~\partial_z \Theta(z) \hspace{5mm}	\mbox{for i = 3, otherwise 0} ~.
\label{reynoldsdiffusivity}
\ee 
The eddy diffusivity $\kappa_{turb}(z)$ is supposed to be a property of the {\em flow}, not of the fluid.  The physical flow properties on which $\kappa_{turb}(z)$ can depend, are the distance $z$ from the plate and the characteristic velocity scale $u_*$ of the turbulent fluctuations, defined by the wall stress $\nu U_{x|z}(z=0) \equiv u_*^2$, the shear rate or drag on the wall. From dimensional reasons we write 
\be
\kappa_{turb} \hat{=} \bar{\kappa}_{\theta} ~z ~u_* ~.
\label{mixinglengthansatz}   
\ee
Here the factor $\bar{\kappa}_{\theta}$ is the dimensionless thermal von K\'arm\'an constant, whose empirical value depends on the flow type and for Rayleigh-B\'enard flow is not known.  

The equation for the temperature profile then is $\partial_z [0 + \kappa_{turb} (z) \partial_z \Theta (z) + \kappa \partial_z \Theta(z)] $. Integrate from $z=0$ to some arbitrary $z$ and use $\kappa_{turb}(z=0) = 0$ to find $(\kappa_{turb} + \kappa)\partial_z\Theta(z) - \kappa\partial_z\Theta(z)|_0 = 0$. The last term is just $J$ or $Nu \cdot \kappa \Delta L^{-1}$, where $\Delta = T_b - T_t$ is the temperature difference between the bottom and the top plates, which causes the thermal flow. $J$ is $z$-independent. This leads to the thermal profile equation
\be
\partial_z \Theta (z) = \frac{-J}{\kappa + \bar{\kappa}_{\theta} ~z ~u_*} ~.
\label{profileequation}
\ee 
If $J$ is positive, i. e., the heat flows upwards, the thermal gradient is negative, $\Theta(z)$ decreases, as is characteristic for the bottom plate. With the choice $\Theta (z=0) = 0$ at the bottom plate we have $\Theta(z) \equiv T(z) - T_b$ relating $\Theta$ with the physical temperature $T(z)$. 

From this equation we draw conclusions for the bottom part profile. In the immediate vicinity of the plate it is $\kappa \gg \kappa_{turb}$, giving rise to the "linear thermal sublayer" of the profile, $\Theta(z) = - J \cdot (z/\kappa)$, $0 \le z \lesssim z_{*,\kappa}$. This linear thermal sublayer extends until $z = z_{*,\kappa}$ for which $\kappa \hat{=} \bar{\kappa_{\theta} ~ u_* ~z_{*,\kappa}}$, i.e., if the molecular and the eddy thermal diffusivity are of the same size. The relation to the kinetic sublayer width $z_*$ is 
\be
z_{*,\kappa} \equiv \frac{\kappa}{\bar{\kappa}_{\theta} u_*}  ~~~\rightarrow ~~~ z_{*,\kappa} = Pr^{-1} \frac{\nu}{\bar{\kappa}_{\theta} u_*} = Pr^{-1} \frac{z_*}{\bar{\kappa}_{\theta}}, ~~~\mbox{with}~~ z_* \equiv \frac{\nu}{u_*} ~. 
\label{linearwidths}
\ee  

Beyond the linear thermal sublayer the profile is increasingly dominated by the eddy diffusivity, $\kappa_{turb} \gg \kappa$, implying $\partial_z \Theta(z) = -J / (\bar{\kappa}_{\theta} u_* z)$ and thus 
\be
\Theta(z) = - \frac{J}{\bar{\kappa}_{\theta} u_*} \left( \mbox{ln} \frac{z}{z_*} + f \right) 
\label{logprofile}
\ee   
is the thermal profile in the turbulent BL. The integration constant $f$ may depend, of course, on $\kappa$ and $\nu$, but only in the form $\nu / \kappa$, since $f$ has unit 1. Thus $f = f(Pr)$, which is reported in \cite{ll87} to empirically be about 1.5 for air, i. e., for $Pr \approx 0.7$.   

To make use of this formula for insight into the thermal log-profile we need the strength of the velocity fluctuations $u_*$ as a function of $Re(Ra)$. This has been derived in \cite{gro11}. But before determining $u_*$ we make now contact to the experimentally measured profile, cf. \cite{ahl12}. The experimental log-profile is parametrized in the form 
\be
\frac{T(z) - T_m}{T_b - T_t} = A ~ \mbox{ln} \frac{z}{L}  + B ~.
\label{experimental-profile}  
\ee                  
Here $T_m = (T_b + T_t)/2 = T_b - \Delta /2$ is the mean temperature in the Rayleigh-B\'enard cell. Comparing eqs.(\ref{logprofile}) and (\ref{experimental-profile}) leads to 
\be
\frac{\Theta(z)}{\Delta} = A ~ \mbox{ln} \frac{z}{L} + B - \frac{1}{2} = - \frac{J}{\bar{\kappa}_{\theta} u_* \Delta} \left( \mbox{ln} \frac{z}{L} + \mbox{ln} \frac{L}{z_*} + f \right)~.
\label{dimensionlessprofile}
\ee
We now can identify the dimensionless empirical parameters $A$ and $B$. 
\be
A \equiv - \frac{\kappa Nu}{L \bar{\kappa}_{\theta} u_*}
\label{A}
\ee
and
\be
B = \frac{1}{2} - \frac{\kappa Nu}{L \bar{\kappa}_{\theta} u_*} \left[ \mbox{ln} \frac{L}{z_*} + f(Pr)  \right] = A \left[ \mbox{ln} \frac{L}{z_*} + f(Pr) \right] + \frac{1}{2} ~.
\label{B}
\ee     
    
One easily verifies that the dimensions of $A$ and $B$ are indeed 1. Besides trivial parameters and an yet undetermined empirical parameter, the thermal von K\'arm\'an constant $\bar{\kappa}_{\theta}$, two physical quantities determine the measured constants $A$ and $B$. This is first the strength of the heat flux $J = Nu \cdot \kappa \Delta L^{-1}$, describing the amplitude of the log-profile, and second the strength of the turbulent velocity fluctuations $u_*$. While $Nu$ and its scaling behavior in the ultimate range is experimentally (and theoretically) rather well known, the velocity amplitude $u_*$ has been calculated in \cite{gro11}. We make use of those results now.

\section{The log-profile parameters $A$ and $B$}

\subsection{Parameter $A$}

We start with the discussion of the amplitude $A$ of the log-profile. From eq.(\ref{A}) the parameter $A$ apparently can be written as 
\be 
A = - \frac{\kappa Nu}{L \bar{\kappa}_{\theta} u_*} = - \frac{Re^{-1} Pr^{-1} Nu}{\bar{\kappa}_{\theta} u_*/U} 
\label{A-2}
\ee
with $Re \equiv U L / \nu$. Here we have introduced the wind amplitude $U$ and the corresponding Reynolds number $Re$. For not yet too large $Ra$ this amplitude $U$ usually is visualized as a large scale circulation (LSC) in the RB sample. For very large $Ra$ on the other hand, which are considered here, such LSC will probably not survive under the strong turbulent fluctuations. But its remnants locally in space and time still must have physical importance, since there apparently is enough shear in the plates' boundary layers to induce transition to turbulence, as it can be observed experimentally in the scaling exponents of the Nusselt number $Nu$ and the Reynolds number $Re$ versus $Ra$ as well as the measured characteristic changes of the scaling exponents indicating this transition, see \cite{fun09, ahl09b, ahl10, he12a, ahl12a} and also \cite{ahl12}.

To quantify this we use results from \cite{he12}. In the ultimate range $Ra \gtrsim 10^{15}$ the effective Reynolds number was found to be $Re = 0.0439 \times Ra^{0.50}$, resulting in $Re(Ra=10^{15})~ = ~1.39\times 10^6$. Extrapolating $Re$ from smaller $Ra$, known as the classical range of RB flow, in \cite{he12} it is found $Re=0.407 Ra^{0.423}$, the scaling exponent being well consistent with the GL theory \cite{gro00}. According to this classical range formula the wind amplitude would be measurably smaller at $Ra=10^{15}$, namely $Re=0.901\times10^6$. 

In the cited RB experiment it is $Nu = 5631$ and $Pr = 0.859$ at $Ra = 1.075 \times 10^{15}$. From this we can calculate the coherence length in the turbulent bulk, which is defined as $\ell_{coh} \approx 10 \eta_{Kol} = 10(\nu^3 / \epsilon)^{1/4} = 10(Pr^{-2} Ra Nu)^{-1/4} L$. One gets $\ell_{coh} /L = 1.87 \times 10^{-4}$ or in physical units (using $L = 2.24$ m for the height of the Goettingen Uboot device) $\ell_{coh} = 0.419$ mm. This sets the lower bound of the turbulence eddy sizes in the bulk. There will be larger eddies too, whose extension is between the external scale of order $L$ ($=D$ in the case $\Gamma=1$) and $\ell_{coh}$. These even if fluctuating temporally and in position can provide the necessary shear in the boundary layers. 

To quantify this we compare with the thickness $z_*$ of the linear viscous sublayer above the plate, which will be introduced and calculated later. It will be estimated as $z_* / L = 1.98 \times 10^{-5}$ for $Ra = 10^{15}$. Then $z_* / \ell_{coh} = 0.106$, meaning that even the smallest eddies of the bulk cascade are still 10 times larger than the kinetic viscous sublayer extension. That holds all the more for the energy carrying larger eddies up to the macroscocic scale $L$. Thus the bulk flow, even if strongly fluctuating, can provide sufficient shear to imply the turbulence transition at the plates (and most probably also at the side walls).              

Having clarified the meaning and physical importance of the wind amplitude $U$, we can determine the log-profile and its parameters $A$ and $B$. In \cite{gro11} we have derived the following two expressions for the Nusselt number $Nu$ and the velocity fluctuation amplitude relative to the given asymptotic flow velocity, $u_* / U$, in the logarithmic ultimate range, cf. eqs. (20) and (2) of that paper, namely
\be
Nu = \frac{\frac{\bar{\kappa}_{\theta}}{2} \frac{u_*}{U} Re Pr}{\mbox{ln} \left( Re \frac{u_*}{U} \frac{1}{b} \right) + \tilde{f}(Pr)}. 
\label{Nu}
\ee 
The fluctuation amplitude $u_*$ solves the equation 
\be
\frac{u_*}{U} = \frac{\bar{\kappa}}{\mbox{ln} \left( Re \frac{u_*}{U} \frac{1}{b} \right) }~.
\label{u}     
\ee
Here $b$ is an empirical parameter of the velocity profile. It characterizes the position of the buffer range, in which the transition occurs from the linear viscous sublayer to the log-layer range in the velocity profile. Usually that profile -- known as the {\em law of the wall} -- is written in the form 
\be
\frac{U_x(z)}{u_*} = \frac{1}{\bar{\kappa}} \mbox{ln} \left( \frac{z u_*}{\nu} \right) + B_u~. 
\label{u-profile} 
\ee
Here $\bar{\kappa}$ is the kinetic von K\'arm\'an constant, taken in the following as $\bar{\kappa} = 0.4$. $B_u$ describes the velocity $U(z)$ at $z = z_*$, i. e., it characterizes the buffer range of the {\em velocity} profile by indicating, how far the log-law velocity $U(z)$ is shifted in amplitude, if extended to the linear viscous sublayer. There are various values of the empirical constant $B_u$ given in the literature, such as e. g. $5.1$, cf. \cite{ll87}, \cite{pop00}. To the best of our knowledge there is not yet any information available for its value in the case of Rayleigh-B\'enard flow in a closed container. Note that this RB-flow neither is stream wise infinite nor does it span wise approach a given constant amplitude $U$, but rather becomes small again near the mid plane and then even changes sign. Thus $B_u$ remains a parameter still to be measured in RB cells. We include it into the profile equation (\ref{u}) in the form 
\be
b \equiv e^{- \bar{\kappa} B_u}~.
\label{b-definition}
\ee  
According to eq.(\ref{u}) the numerical value of $b$ determines the amplitude of the velocity fluctuation scale $u_*$ (or shear stress $\propto u_*^2$) together with the Reynolds number $Re$ in the combination $Re / b$. From (\ref{u}) we see that the smaller  $b$ is the smaller $u_*$ will be. Similarly $u_*$ will decrease with increasing Reynolds number $Re$ and thus with growing Rayleigh number $Ra$. For $Re(Ra)$ we meanwhile have experimental information, see \cite{he12}: In the ultimate state it is $Re_{eff} = 0.0439 Ra^{0.50}$. We thus have typical $Re$-numbers of order $10^6$ or more in the ultimate range of thermal convection with $Ra \sim 10^{15}$. 

The velocity profile parameter $b$ also enters the $Pr$-dependence of the temperature profile via the definition $\tilde{f}(Pr) = f(Pr) + \mbox{ln} b/2$. Using all these previous results we can express the amplitude parameter $A$ in the form 
\be 
A = - \frac{1}{2 \left[ \mbox{ln} \left( \frac{u_*}{U} Re \frac{1}{b} \right) + \tilde{f}(Pr)  \right]} = - \frac{1}{2 \left[\bar{\kappa} \frac{U}{u_*} + \tilde{f}(Pr)   \right]} ~. 
\label{A-formula}
\ee  
Two remarks may be useful. In this expression for $A$ the thermal von K\'arm\'an constant $\bar{\kappa}_{\theta}$ does not appear explicitly; instead the kinetic one, $\bar{\kappa}$, shows up. This comes from eliminating $Nu / \bar{\kappa}_{\theta}$ in the defining equation (\ref{A-2}) for $A$ by using eq.(\ref{Nu}). Second, the originally derived explicit $Nu$-dependence can be completely substituted and expressed in terms of the log-corrections originating from the log-profile of the velocity. The only reminder to the thermal profile is the thermal buffer range shift $\tilde{f}(Pr)$, the thermal analog of the corresponding kinetic shift $B_u$. 

Both $A$ and $B$ depend on $Re$ as well as on $u_* /U$, and thus on $Ra$. This implies the $Ra$-dependence of these measured fit-amplitudes $A$ and $B$. Numerical values for $Re$ and $u_* /U$ have been given in Table I of reference \cite{gro11} for $Ra = 10^{14}$ and $b=1$. For any other $Ra$ (and $b$) they can be computed by solving the implicit equation (\ref{u}), either numerically or via its continued fraction representation given in \cite{gro11}. We do this below for $Ra = 10^{15}$.  

Taking for the parameters $f$ and $b$ the values reported in textbooks for channel or pipe flow of gases, i. e., $f(Pr=0.7)\approx 1.5$ (cf. \cite{ll87}) and $b$ = O(1), one finds $\tilde{f}(Pr)$ = O(1). The estimate for the magnitude of the relative fluctuations $u_* / U$ in \cite{gro11} gave values of the order of 0.05. Thus the first term in the denominator of eq. (\ref{A-formula}) is an order of magnitude larger, even more so for larger $Re(Ra)$. Neglecting therefore $\tilde{f}$ we find the following approximate formula for the parameter $A$:
\be
A \approx - \frac{1}{2 \mbox{ln} \left( \frac{u_*}{U} Re \frac{1}{b}  \right)}  ~ = ~ - \frac{1}{2 \bar{\kappa}} \frac{u_*}{U} ~.
\label{A-approximate}      
\ee
For $Ra = 10^{14}$ it is $u_* / U = 0.039$ and $0.065$ for the choices $Re_1$ and $Re_2$ in table I of \cite{gro11}, respectively, leading to the numerical values $A = -0.0488$ and $A = - 0.0813$, respectively. For $Ra=10^{15}$ and $Re=1.4\times10^6$ we calculate from eq.(\ref{u}) (and the choice $b=1$) the value $u_* / U = 0.036$, leading to $A=-0.045$, still about half an order of magnitude larger as the measured value $A_{exp} = 0.0082$, given in \cite{ahl12}. 

The scaling of the profile parameters $A$ and $B$ with the Rayleigh number $Ra$ can be determined as follows. First transform the $Re$-dependence of $u_*/U$ into a scaling Ansatz, $u_*/U \propto Re^{\alpha'}$. Here $\alpha' = \alpha'(Re)$ is a local exponent, decreasing with $Re$ and therefore with $Ra$. In the relevant $Re$-range one calculates $\alpha' = -0.0855$. The negative scaling exponent $\alpha'$ means that the ratio $u_*/U = Re_*/Re$ {\em decreases} with increasing $Re$ and corresponding increase of $Ra$. But, of course, the turbulent fluctuation scale $Re_* = u_* L / \nu$ itself increases in size for stronger thermal driving $Ra$. It is 
\be
Re_* \propto Re^{1+\alpha} = Re^{0.9145} ~~, ~~~\mbox{in the $Ra$-range $10^{15}$} ~. 
\label{Re-scaling}
\ee  
Introducing now this $Re$-scaling of $Re_*$ into eq.(\ref{A-approximate}) results in 
\be
A \propto Re^{\alpha'} \propto Ra^{0.50 \alpha'} = Ra^{-0.043} ~~, ~~~\mbox{in the $Ra$-range $10^{15}$} ~.  
\label{Ra-scaling}
\ee
The amplitude $A$ thus decreases with $Ra$, though very slowly. Over the next decade it will be smaller by approximately a factor of 0.906, i.e., its value at $Ra$ is expected to be about 9.4 \% less. This predicted slow decrease of $A$ with $Ra$ could be consistent with experimental observation \cite{ahl12}. We emphasize that $\alpha'$ decreases even further with increasing $Re$, i.e., asymptotially in $Ra$ the amplitude $A$ will approach a constant, $Ra$-independent limit $A_{\infty}$. 

One may wish to try coming closer to the experimental value for $A$ in the measured range of $Re \approx 10^6$ and $Ra = 10^{15}$ (\cite{he12}) by adjusting the still badly known value of $b$. Instead of $b=$O(1) one could reduce it, mimicing thus a larger $Re$, which leads to smaller $u_*/U$. To be more precise: Take the measured $u_* /U = 0.0082$ and $Re(Ra=10^{15}) = 1.4 \times 10^6$ and insert this into (\ref{u}). This then determines $b$; we obtain ln$(1/b) = \bar{\kappa} B_u \approx 40$, thus $B_u \approx 100$, and $b$ close to 0, namely $b = \mbox{O}(10^{-17})$. 

Such unusually large value of $B_u$ indicates that the model for the velocity field's $U$-profile is indeed oversimplified. An important missing feature when assuming an asymptotically constant flow amplitude $U$ with increasing distance $z$ from the plate is as follows. The LSC in an RB-sample does {\em not} approach a nonzero constant value when going off the plate. Instead $U(z)$ not only will decrease again, it even passes through $0$ at $z/L = 0.5$ changing sign in the upper half of the RB cell. Thus between about $0.25 \lesssim z/L \lesssim 0.5$ the average flow has a negative instead of a positive or zero $z$-slope. The thermal flux $J$ keeps its positive sign. Because of this quite different $U(z)$ profile in comparison to pipe or channel flow we have to expect a deviation from the temperature log-profile in the range above $z/L \gtrsim 0.25$. Indeed such deviation can be seen in experiment, see \cite{ahl12}, Fig.1. -- Another consequence of such deviation from the p!
 ure log-profile is that the intimate connection between the amplitudes $A$ and $B$, to be discussed in the next chapter, will loose validity.

\subsection{Parameter $B$}

Let us now analyze the parameter $B$ as derived in eq. (\ref{B}), in particular the right hand part of this equation. Consider first the term $\mbox{ln} \frac{L}{z_*}$. With the above given definition $z_* = \nu / u_*$ one obtains $L/z_* = L u_* /\nu $ and thus $B = A \left[ \mbox{ln} \left( Re \frac{u_*}{U} \right) + f(Pr) \right] + 1/2 = A \left[ \mbox{ln} \left( Re \frac{u_*}{U} \frac{1}{b} \right) + \tilde{f}(Pr) + \mbox{ln} 2 \right] + 1/2$. Now, the ln-term plus the next one are just $- \frac{1}{2 A}$, which after multiplication with $A$ leaves $-1/2$, canceling the final term 1/2. Thus 
\be
B = A \cdot \mbox{ln} 2 ~~~ \mbox{or} ~~~ B/A = \mbox{ln} 2 = 0.693... ~.
\label{B-A}
\ee
Irrespective of any approximate calculation of the coefficient $A$, the other coefficient $B$ is always about $70 \%$ of $A$, in particular is also negative. Using this the $T$-profile can be written as 
\be
\frac{T(z) - T_m}{T_b - T_t} = A \left( \mbox{ln} \frac{z}{L} + \mbox{ln} 2 \right) = A ~ \mbox{ln} \frac{z}{L/2} ~.  
\label{profile-2}
\ee 
This is consistent with the underlying idea that $T(z = L/2) = T_m$, which has been used when relating the $T$-profile, expressed in terms of the thermal current $J$ or $Nu$, with the driving temperature difference $\Delta = T_b - T_t$, and which also is an immediate consequence of the Ansatz (\ref{experimental-profile}) for fitting the data.

Let us come back to the consequences of the deviation of the ($x$-component of the) velocity profile from the classical case with asymptotically constant amplitude $U$. In an RB cell the $U(z)$-profile does not stay asymptotically constant with $z$ but goes through a maximum, then decreases further and even changes sign at $z/L \approx 0.5$. This different maximum-and-beyond behavior of the wind leads to deviations in the thermal profile from a log-law. Therefore the intimate relation between $A$ and $B$, viz. $B/A = \mbox{ln} 2$, typical for the log-profile, will no longer be valid. Also this statement is consistent with experimental observation, cf. \cite{ahl12}: the ratio $B/A$ has quite some scatter, and although not too far way it apparently differs from ln$2$.  

To briefly summarize, the thermal profile is characterized by -- starting from the plate -- (i) a tiny linear thermal sublayer of extension an oder of magnitude less than the bulk coherence length, (ii) a buffer range in which the linear increase in the sublayer turns over into the (iii) log-law, observable over a broader $z$-range up to about a quarter of the RB cell height, then (iv) changing the profile again due to the decrease of the blowing wind including its directional change, from positive to negative (or vice versa), which may be denoted as the temperature's center profile. This latter one, the center part, still has to be explored in more detail.

\section{Position dependence of profile parameters $A(r)$ and $B(r)$}

The basic assumption of the just given derivation of the profile parameters $A, B$ is a plane parallel homogeneous flow with veloctity $U$ over an infinitely extended plate. In a Rayleigh-B\'enard cell this -at best- is realized in the center-range of the circular cell, i. e., at $r = 0$. We now make a crude model for the shape of the large scale circulation (LSC). We have in mind the case of aspect ratio $\Gamma = 1$, but one can argue similarly for $\Gamma = 1/2$ (and other). Then, if one moves away from the center range at $r=0$, the relevant flow velocity leading to the velocity shear in the BL and the perpendicular logarithmic profiles of velocity and temperature is the $x$-component of the LSC only. We therefore have to substitute in above formulas always $U \Rightarrow U_x = U_x(r)$. If we consider for simplicity a circular LSC, we get $U_x = U \mbox{cos} \phi = U \sqrt{1 - \frac{r^2}{R^2}}$; here $U$ still denotes the LSC amplitude, which defines $Re=UL/\nu$.

Considering expression (\ref{A-approximate}) will result in an $r$-dependent profile coefficient 
\be
A(r) = \frac{A}{\sqrt{1 - (r/R)^2}}~, ~~~ \mbox{with $A$ the coefficient at the center}. 
\label{A(r)}        
\ee
Expressed in terms of the (relative) wall distance $\xi \equiv \frac{R-r}{R}$, thus $0 \le \xi \le 1$ between the wall and the center, respectively, it is 
\be
A(\xi) = \frac{A}{\sqrt{2\xi - \xi^2}} \approx \frac{A}{\sqrt 2} \frac{1}{\xi^{1/2}}. 
\label{A(xi)}
\ee
The coefficient $A(\xi)$ decreases with increasing distance from the wall $\propto \xi^{-1/2}$. This result is consistent with the experimental finding of an $A$-decrease towards the center. The data presented in \cite{ahl12} have been taken at the position $(R-r)/L = 0.0045$, corresponding to $\xi = 0.009$. This implies $A(\frac{r}{R} = 0.991) = A(\xi = 0.009) = 7.47 A$. 

We note that these last formulas cease to be valid in the limit $\xi \rightarrow 0$ or $r \rightarrow R$, because for sufficiently small $U_x(r)$ the BL locally is no longer turbulent, finally there is only upward flow and one is in the BL of the side wall. But note further that with the full, non-approximate expression (\ref{A-formula}) for the profile coefficient $A$ the additional term in the denominator weakens the $r$-dependence, all the more the larger $r$ becomes. Then even the limit $r \rightarrow R$ exists, leading to 
\be
A(r \rightarrow R) = - \frac{1}{2 \tilde{f}(Pr)} ~.
\label{A-limit}
\ee      

One can draw various conclusions from the r-dependence of A(r) on the shape of the LSC as follows. Note first that the magnitude of the LSC wind is everywhere the same along its closed trajectory, called $U$ or in dimensionless form $Re$. The direction of the wind varies along the LSC orbit. Its local unit vector be called $\vec{t}(s)$, which is the  local tangential unit vector along the (closed) LSC curve. $s$ denotes the arc length along the LSC curve. Then the velocity relevant for $A(r)$ is given by $U_x = \vec{e_x} \cdot \vec{t}~U$, the x-component of the local LSC velocity vector $U\vec{t}$.  Given a model for the LSC-curve, e. g. a circle or an ellipse, this can be described in terms of a suitable parameter $\phi$ by $\vec{x}(\phi)$. Expressing the (arbitrarily given) parameter $\phi$ in terms of the arc length $s$, the tangential vector then is the derivative of the curve, i.e., $\vec{t} = \frac{d \vec{x}(s)}{ds}$. Measuring on the other hand $U_x(r)$ via $A(r)$ all!
 ows to re-construct the LSC curve.

Some qualitative features of the r-dependence of $A(r)$ are the following: (i) For small $r$ the amplitude $A(r)$ always increases as $A(r) = A (1 + \mbox{const} (\frac{r}{R})^2)$; there is no term linear in $r$ because of analyticity reasons. (For a strictly circular flow it is $\mbox{const} = 0.5$). (ii) For $r$ near the side wall, i. e., for small $\xi = (R-r)/R$, the amplitude $A(\xi) \sim \xi^{-n}$ decreases with increasing distance from the wall with an exponent $n= 1/2$  for a circular LSC; for an elliptically shaped wind curve it will be steeper, thus $n$ is larger. In the particular case of aspect ratio $\Gamma = 1 / 2$ there may be two circular rolls above each other, which again would lead to $n = 1 / 2$. In case there is only one single roll this will be elliptically shaped and thus $n$ is larger. In case both LSC shapes are present part of the time, the exponent $n$ will be somewhere in between, i. e., in any case one would find a steeper decrease of A with the !
 wall distance as compared to the circular case for $\Gamma = 1$. This is consistent with the measurement of  $n\approx 2/3$ in the case $\Gamma = 0.5$ (Goettingen Uboot team, private communication). (iii) If $\Gamma$ is a little below 1, $n$ will be a bit larger than $1/2$; if $\Gamma$ is a little above 1, $n$ is expected to be somewhat smaller than $1/2$; etc.

\section{Concluding remarks}

We have calculated the profile parameters $A$ and $B$, defined in eq.(\ref{experimental-profile}), of the experimentally measured logarithmic temperature profiles in the ultimate state of Rayleigh-B\'enard convection as realized for very large $Ra$. In the case of a pure thermal log-law as well as reflection symmetry with respect to the middle plane $z = L/2 = R$ of the $\Gamma = 1$ cylindrical sample, which we expect and which should hold for approximately temperature independent material properties, the coefficient $B$ equals $A$ up to a factor ln 2. Since the real wind profile is  different from the standard case of approachig a constant when going off the plate, in the RB cell instead going down towards the center, even followed by a directional inversion of the wind direction, one finds deviations from this value. 

The amplitude $A$ physically measures the strength of the turbulent velocity fluctuation scale $u_*$ relative to the LSC velocity $U$ (together with the von K\'arm\'an constant $\bar{\kappa}$); if one is sufficiently near to the side wall, it also measures the Prandtl number dependent temperature profile shift constant $\tilde{f}(Pr)$. We explain the $r$-dependence of $A(r)$ by the decreasing magnitude of the local flow velocity $U_x(r)$ parallel to the bottom plate with increasing distance $r$ from the center, which by (\ref{A-approximate}) or (\ref{A-formula}) leads to an increase of the $A$-amplitude with increasing $r$.  

The numerically obtained value for $A$ for the case $b = \mbox{O}(1)$ does not coincide too well with the measured one, if we use the values of the fluctuation scale $u_*$ calculated in \cite{gro11}. When we published that work the meanwhile measured values for the sample's LSC-response $Re = Re(Ra)$ (cf. \cite{he12}) was not yet known and assumed to be smaller; if our interpretation is correct it means, that the experimental $u_* / U$ values are smaller in an RB-sample, because $U \propto Re$ is larger than assumed previously and $b$ is smaller.        

Thus there is very interesting information in the measured $T$-profile parameters $A(r)$ and $B(r)$. But, as usual in Rayleigh-B\'enard flow, to complete, check and confirm theoretical interpretation and explanation will need information also about the velocity profile and its expected considerable deviations from the model of classical channel or pipe flow as is used here for a lack of better.

\vspace{0.5cm}
\noindent \textbf{Acknowledgements:} We very much thank Guenter Ahlers for various stimulating discussions.

\end{document}